\documentclass[reprint, floatfix, amsmath, amssymb, aps, prl, showkeys, showpacs]{revtex4-1}

\usepackage{graphicx}
\usepackage{bm}
\usepackage{hyperref}
\usepackage{color}
\usepackage{pgfplots}

\pgfplotsset{compat=1.13}
\pgfplotsset{every axis/.append style={font=\footnotesize}}

\newcommand{\vecs}{\hat{s}}
\newcommand{\vecr}{\vec{r}}

\newcommand{\veck}{\vec{k}}
\newcommand{\vecm}{\vec{m}}
\newcommand{\mua}{\mu_a}
\newcommand{\mut}{\mu_t}
\newcommand{\mus}{\mu_s}

\newcommand{\ddt}{\frac{\partial}{\partial t}}
\newcommand{\urs}{\phi(\vecr, \vecs, t)}
\newcommand{\ums}{\phi(\vecm, \vecs, t)}

\newcommand{\ursp}{\phi(\vecr, \vecs', t)}
\newcommand{\qrs}{q(\vecr, \vecs, t)}
\newcommand{\fpsilm}{\hat{\psi}_\ell^m}
\newcommand{\dfpsilm}{\dot{\psi}_\ell^m}

\newcommand{\snm}{{\mathcal{S}^{N-1}}}

\newcommand{\drm}{\mathrm{d}}
\newcommand{\abs}[1]{\left|#1\right|}
\newcommand{\afn}[2]{\alpha_{#1}^{#2}}
\newcommand{\bfn}[2]{\beta_{#1}^{#2}}
\newcommand{\im}{\mathrm{i}\mkern1mu}

\newcommand{\Ylm}{Y_{\ell}^{m}}
\newcommand{\sphsum}{\sum_{\ell=0}^{\infty} \sum_{m={-\ell}}^{\ell}  }

\renewcommand{\vec}[1]{\mathbf{#1}}

\let\oldhat\hat
\renewcommand{\hat}[1]{\oldhat{\mathbf{#1}}}

\begin{document}

\title{A pseudospectral method for solution of the radiative transport equation}

\author{Samuel Powell}
\affiliation{
School of Biomedical Engineering \& Imaging Sciences
\\King's College London, Becket House, London, SE1 7EU, UK.}

\author{Ben~T.~Cox}
\affiliation{
Department of Medical Physics \& Biomedical Engineering
\\University College London, London, WC1E 6BT, UK.}

\author{Simon~R.~Arridge}
\affiliation{
Centre for Medical Image Computing
\\ University College London, London, WC1E 6BT, UK.}

\date{\today}

\begin{abstract}
The radiative transport equation accurately describes light transport in participating media, though analytic solutions are known only for simple geometries. We present a pseudospectral technique to efficiently compute numerical solutions to large-scale time-dependent and steady-state problems, with anisotropic scattering. A perfectly matched layer is proposed which allows this method to be applied in arbitrarily complex heterogeneous media. Our GPU-accelerated implementation of the technique is validated by comparison with a Monte-Carlo simulation, demonstrating excellent agreement.
\end{abstract}

\maketitle

\section{\label{sec:intro}Introduction}

The radiative transport equation (RTE) is a mono-energetic form of the Boltzmann transport equation, which models the propagation, attenuation and scattering of particles within participating media \cite{case1967linear}, and has been applied in numerous fields, including neutron transport, atmospheric physics. In this work we consider the application of the RTE in the field of biomedical optics\cite{Chandrasekhar:1511024, Ishimaru:1999wx}, where steady-state and time-domain solutions are sought to verify experiments and as part of model-based inverse problems.

The RTE, which can be derived directly from Maxwell's equations \cite{Mishchenko:2006ic}, describes the distribution of specific intensity, or \emph{radiance} within a domain $\Omega$,
\begin{multline}
\left (\frac{1}{c} \ddt + \vecs \cdot\nabla + \mut(\vecr) \right) \urs \\
= \mus(\vecr) \int_{\snm} p(\vecs \cdot \vecs') \ursp \drm \vecs' + \qrs
\label{eq:rte}
\end{multline}
where the change in radiance $\urs$ at a point $\vec{r} \in \Omega$ in direction $\vecs$ at time $t$, is given as a balance of energy lost from an attenuation term $\mu_t = \mu_a + \mu_s$ accounting for absorption ($\mu_a$) and out-scattering ($\mu_s$), inwards scattering from $\vec{s}'$ to $\vec{s}$, and any sources $\qrs$. In the biomedical imaging modalities of interest here, the Henyey-Greenstein phase function \cite{Henyey:1941wt} is typically employed to describe the anisotropic scattering process,
\begin{equation}
p(\vecs \cdot \vecs')  = 
 \frac{1}{4\pi}
 \frac{1-g^2}{(1+g^2-2g (\vecs\cdot\vecs'))^{3/2}}.
 \label{eq:hgfun}
\end{equation}
Boundary conditions for the RTE specify that $\ums = 0$ for $\vecs \cdot \vec{\hat{n}} < 0$, where $\vec{\hat{n}}$ is the outward normal to the boundary of the domain at $\vecm \in \partial\Omega$: that is to say that in the absence of sources, there is no component of the radiance inwards across the boundary.

The high dimensionality of the field variable $\urs \in \mathcal{R}^N \times \mathcal{S}^{N-1}$, and the presence of non-local spatial and angular operators, is such that analytically solutions to the RTE have been found only for homogeneous infinite and semi-infinite geometries\cite{Liemert:2013ky}, and layered media\cite{Liemert:2017jh}. 

Numerical solutions to the RTE in arbitrary three-dimensional geometries are typically sought by stochastic Monte-Carlo (MC) estimates. This approach converges to the solution in the limit of an infinite number of trials, and is inherently parallelisable. Despite the availability of a number of high performance codes, it remains a challenge to generate high quality estimates at distances far from a source, where statistical noise dominates. Furthermore, practical implementations require the use of variance reduction techniques which produce complex geometry dependent noise statistics which prevents modelling of the noise statistics as part of, e.g., an image reconstruction procedure.

Deterministic methods can be categorised by how they treat the angular discretisation.
\begin{itemize}
    \item Discrete ordinate ($S_N$) methods discretise the angular basis over a set of vectors equally spaced over the unit-circle or sphere. $S_N$ methods suffer from ray effects and may require stabilisation which affects both the performance and accuracy of their solutions.
    \item Spherical harmonic discretisations ($P_N$ methods) expand the solution in the natural orthogonal basis over the sphere, and truncate the series to allow numerical solution. $P_N$ methods can suffer from wave-like distortion owing to the truncation of the angular basis, but perform well in highly scattering media such as biological tissue.
\end{itemize}
The use of the RTE within biomedical optics extends across a number of modalities. Low-resolution time-resolved modalities such as diffuse optical tomography (DOT) require time-dependent solutions on picosecond temporal scales in domains c. 100mm on a side, with feature sizes of a few millimeters. High-resolution hybrid modalities such as photoacoustic tomography require steady-state solutions in smaller domains c. 10mm on a side, but may require sub-millimeter resolution. In all cases, domains are scattering dominated, with scattering coefficients c 10mm$^{-1}$ compared to absoprtion coefficeints of c. 0.01mm$^{-1}$, at the near-infrared wavelengths of interest. In this diffuse regime, $P_N$ methods are particularly appropriate as the smooth radiance fields which result are accurately captured in approximations of a lower order than would be required in an $S_N$ method.

Discretisation in space can be achieved with the finite-element method \cite{SuryaMohan:2011vn, Tarvainen:2005gh, Badri:2018fo}, though the large size of the resultant system matrices has limited the practical application of this method. Alternatively, finite-difference \cite{Sakami:2002jc} and finite-volume \cite{Asllanaj:2012bz} methods have been proposed, but the inherent derivative operations require fine discretisation and the use of higher order derivative stencils, leading to limitations in their application in large domains. Mesh-free techniques have also been proposed \cite{Kindelan:2010cn}. There exists a significant body of work on solutions to the transport equation in other fields, such as neutron transport \cite{Wareing:2001aa}. Here $S_N$ methods are more appropriate, and computational performance requirements can be relaxed since the models are not typically employed as part of iterative inverse solvers.

The lack of accurate and computationally efficient deterministic methods often leads to the application of the diffusion approximation (DA) to the RTE. The DA fails in regions of low scattering, and where the gradient of the energy density is large. Furthermore, the DA is parabolic (non-causal) in contrast with the hyperbolic RTE from which it is derived. Hybrid transport-diffusion methods have been proposed to balance the computational challenges of transport solutions with the approximations inherent in the DA \cite{Densmore:2012kx, Roger:2014bu}.

Whilst these differences have until recently been tolerated in the context of diffuse optics, new developments in high-density, time-domain, and hybrid diffuse optical methods require the use of the RTE to achieve their full potential.

\section{\label{sec:pstransport}The pseudospectral method}

Pseudo-spectral methods are typically applied to solve partial differential equations which contain point-wise potential operators. The technique exploits the $\mathcal{O}(N \log N)$ performance of the Fast Fourier Transform (FFT) to take spatial derivatives in the spectral domain (k-space) before applying an inverse transform to apply the potential operators in Cartesian space: thus, all operators are applied in the space in which they are diagonalised, and a far coarser spatial discretisation can be employed, limited only by the Nyquist criterion. Such techniques thus require far less memory and computational effort than their finite-difference or finite-element counterparts.

Pseudo-spectral schemes can be constructed for the RTE using both the $S_N$ and $P_N$ methods, however the elegance of our proposed technique is to exploit the fact that the scattering operator of the RTE is also diagonalised in the $P_N$ approximation, and that directional derivative term forms a recurrence relation over the spherical harmonics which is trivial to compute. We thus operate solely in the angular-spectral domain.

To derive our method we expand the radiance and source terms in a Spherical Harmonic basis,
\begin{equation}
    \urs = \sphsum \sqrt{\frac{2\ell + 1}{4\pi}} \psi_{\ell}^{m}(\vecr, t) \Ylm(\vecs),
    \label{eq:usph}
\end{equation}
\begin{equation}
   \qrs = \sphsum \sqrt{\frac{2\ell + 1}{4\pi}} 
   q_{\ell}^{m}(\vecr, t) \Ylm(\vecs), 
   \label{eq:qsph}
\end{equation}
and employ the addition theorem for spherical harmonics to restate the phase function,
\begin{equation}
    p(\vecs \cdot \vecs') = \sphsum g^{\ell} \bar{Y}_{\ell}^{m}(\vecs') \Ylm(\vecs).
    \label{eq:psph}
\end{equation}
Equations \ref{eq:usph} through \ref{eq:psph} are substituted into equation \ref{eq:rte}, and the result is projected into the spherical harmonic basis by taking the inner product with $\bar{Y}_{\ell}^{m}(\vecs)$. Following some algebra, this yields a set of coupled first order PDEs \cite{Arridge:1999kd},
\begin{multline}
\left ( \frac{1}{c} \ddt + \mut \right ) \psi_{\ell}^{m}  
+\frac{1}{2\ell+1} \left ( \frac{\partial}{\partial z} \left [\afn{\ell+1}{m} \psi_{l+1}^m + \afn{\ell}{m} \psi_{\ell-1}^m \right] \right .\\
-\frac{1}{2} \left ( \frac{\partial}{\partial x} - \im \frac{\partial}{\partial y} \right ) \left [\bfn{\ell}{m} \psi_{\ell-1}^{m-1}  - \bfn{\ell+1}{-m+1} \psi_{\ell+1}^{m-1} \right ]\\
\left . -\frac{1}{2}\left ( \frac{\partial}{\partial x} + \im \frac{\partial}{\partial y} \right ) \left[ - \bfn{\ell}{-m} \psi_{\ell-1}^{m+1} + \bfn{\ell+1}{m+1} \psi_{\ell+1}^{m+1} \right ] \right ) \\
 = \mu_s g^\ell \psi_{\ell}^{m} + q_{\ell}^m,
 \label{eq:coupledr}
\end{multline}
where the terms $\afn{\ell}{m} = \sqrt{(l+m)(l-m)}$ and $\bfn{\ell}{m} = \sqrt{(l+m)(l+m-1)}$ are related to the Clebsch-Gordan coefficients \cite{SuryaMohan:2011vn,Arridge:1999kd}.

To implement our pseudo-spectral scheme we perform an $N$ dimensional spatial Fourier transform of the radiance distribution,
\begin{multline}
    \mathcal{F}[\urs](\vec{k}, \vecs, t) = u(\vec{k}, \vecs, t) \\ 
    = \sphsum \sqrt{\frac{2\ell + 1}{4\pi}}  \Ylm(\vecs) \fpsilm(\veck, t) 
    \label{eq:coupledps}
\end{multline}
where
\begin{equation}
    \fpsilm(\veck, t)  = \int \psi_{\ell}^{m}(\vecr, t) \exp(-\im\vec{k} \cdot \vecr)\; \drm \vecr,
\end{equation}
and $\vec{k} = \{ k_x, k_y, k_z \}$ is the k-space wave-vector. Inserting this result into equation \ref{eq:coupledr} results in a set of coupled ODEs,
\begin{equation}
\left ( \frac{1}{c} \ddt + \mut \right ) \psi_{\ell}^{m} + \mathcal{F}^{-1}[\dfpsilm] = \mu_s g^\ell \psi_{\ell}^{m} + q_{\ell}^m.
\end{equation}
where $\mathcal{F}^{-1}$ denotes the inverse spatial Fourier transform, and
\begin{multline}
\dfpsilm =  \frac{1}{2\ell+1}
\left ( -\im k_z \left  [\afn{\ell+1}{m} \hat{\psi}_{l+1}^m + \afn{\ell}{m} \hat{\psi}_{\ell-1}^m \right] \right . \\
-\frac{1}{2} \left ( -\im k_x - k_y \right ) \left [\bfn{\ell}{m} \hat{\psi}_{\ell-1}^{m-1}  - \bfn{\ell+1}{-m+1} \hat{\psi}_{\ell+1}^{m-1} \right ]\\
\left . -\frac{1}{2}\left (  -\im k_x +  k_y  \right )  \left[ - \bfn{\ell}{-m} \hat{\psi}_{\ell-1}^{m+1} + \bfn{\ell+1}{m+1} \hat{\psi}_{\ell+1}^{m+1} \right ] \right ).
\end{multline}

\subsection{Discretisation, truncation, transient and steady-state solution}

To solve the RTE numerically we form a $P_N$ approximation by closing the recurrence under the assumption that $\psi_{\ell}^m = 0$ for $l>N$, leading to $(N+1)^2$ coupled first-order ODEs. The radiance distribution is sampled in space and the continuous Fourier transform replaced with its discrete counterpart. Finally, we sample in the temporal domain and implement a numerical approximation to the temporal derivative. For brevity, we choose here to employ a first order explicit time-stepping scheme,
\begin{equation}
\psi_{\ell, t+1}^{m} = \psi_{\ell, t}^{m} + c \Delta t \left( \mathcal{F}^{-1}[\dot{\psi}_{\ell,t}^m] + (\mut - \mu_s g^\ell) \psi_{\ell,t}^{m}  \right ) + q_{\ell,t}^m.
\label{eq:dtpsrte}
\end{equation}

The use of an explicit temporal discretisation provides conditional stability, requiring that a suitable choice of time-step $\Delta t$ be employed. We have found that time-steps of the order of 100fs provide stability in the explicit time-stepping scheme across a wide range of physiologically relevant parameters. Typical time-domain measurements acquired using time-correlated single photon counting (TCSPC) hardware provide measurements with time-bins of c. 800fs. Thus, for time-domain measurements the use of an explicit scheme offers reasonable efficiency.

To compute a steady-state solution, this time-domain solution can be numerically integrated, though the small step-size required for stability of the explicit scheme makes this approach inefficient. Alternatively, the time-domain form can be integrated analytically, whence the pseudo-spectral method provides an implicit operator representing the underlying transport equation. Krylov subspace iterative methods can then be employed to find a steady-state solution through repeated application of the implicit operator. This technique is related to the method of exponential integrators, a topic to which we shall return in the discussion.

\subsection{Boundary conditions}

The primary limitation of pseudo-spectral methods is that they imply periodic boundary conditions (the exact form of which depends upon the choice of transform). In many applications a-periodic boundary conditions are enforced by the development of a \emph{perfectly matched layer} (PML), which attenuates outgoing radiation before it reaches the boundary of the \emph{computational} domain. Since the transport operator in the RTE is advective (that is to say that the action of transport on functions in the domain is to translate only along the directions $\vecs$), a PML can be implemented by the simple addition of an additional isotropic loss term outside of the real domain. 
Further, to adhere to the specified boundary conditions, which state that there is no radiation inwards across the boundary of the domain, we must set the scattering coefficient $\mu_s = 0$ in those regions which do not correspond the real domain interest. A full analysis of such boundary conditions was recently analysed in \cite{Egger:2018ti}. We thus solve,
\begin{equation}
\begin{array}{lr}
    \left ( \frac{1}{c} \ddt + \mut \right ) \psi_{\ell}^{m} + \mathcal{F}^{-1}[\dfpsilm] = \mu_s g^\ell \psi_{\ell}^{m} + q_{\ell}^m &\vecr \in \Omega,\\
    \left ( \frac{1}{c} \ddt + \mu_b \right ) \psi_{\ell}^{m} + \mathcal{F}^{-1}[\dfpsilm] = 0 & \vecr \in \Omega_c \setminus \Omega,
\end{array}
\end{equation}
where $\mu_b$ is the PML attenuation term, $\Omega_c$ is the full computational domain, and $\Omega$ is the real domain of interest. The exact spatial distribution of $\mu_b$ should be chosen to ensure that the solution is attenuated to a sufficient degree on the boundary of $\Omega_c$. Choice of $\mu_b$ is straightforward since within $\Omega_c$ the  radiance decays at a rate rate $\exp(-\mu_b \abs{r_b} )$, where $r_b$ is the distance from the boundary in a given direction. It is also desirable for numerical reasons to choose a distribution which leads to a smooth decay of the radiance, in order to limit the introduction of higher spatial frequencies in the spectral representation.

\section{\label{sec:results}Results}

The pseudospectral algorithm was implemented in the Julia programming language \cite{Bezanson:2014tm}. The application of the transport operator occurrs in three stages. 
  \begin{enumerate}
    \item A three-dimensional FFT is applied over the spatial dimensions of the complex valued input radiance field. The directional gradient operator ($\mathbf{s} \cdot \nabla$) is then applied to the resulting k-space coefficients according to eq. (10). The solution is returned from k-space through application of an inverse transform.
    \item The potential operators ($\mu_t - \mu_s g^{\ell}$) are applied to the input radiance field, and the result from (1) is added. This is a simple pointwise operation, and concludes application of the transport operator.
    \item If required, time stepping is applied according to eq. (11) which updates the input radiance field in-place.
  \end{enumerate}
The algorithm defaults to running on the host CPU, where the FFTW library is used to implement the FFT. To improve execution speed we have also implemented the algorithm as a set of custom CUDA kernels which execute on the GPU, in this configuration the cuFFT library is used to implement the FFT.

All simulations were executed on an nVidia Tesla K40 GPU driven by a 32-core Xeon workstation. In the numerical demonstrations which follow, we in each case compare the results of our pseudo-spectral method with a Radiance Monte-Carlo technique we developed previously \cite{Hochuli:2015et, Powell:2017fw}.

\subsection{Transient solution without PML}

The basic algorithm can be validated without the use of the PML by evaluating the time-domain internal radiance distribution over a period before the radiance distribution has propagated to the boundary.

We consider a cube $50$mm on a side of background parameters $\mua = 0.01$mm$^{-1}$, $\mus = 1.0$mm$^{-1}$, and an anisotropy factor of $g=0.9$. The region $x \leq 15$mm has a higher scattering coefficient  $\mus = 5.0$mm$^{-1}$. The domain is discretised into a $64\times64\times64$ grid of points, and the pseudo-spectral method is used to solve the system under a 21-degree spherical harmonic ($P_{21}$) approximation. 
\begin{figure}[ht!]

    \centering
    
    \input{./figures/inf_x.tex}
    \caption{Fluence along the line $y=z=0$mm at three points in time, pseudo-spectral method (solid) vs. Monte-Carlo (dots).}
    \label{fig:inf_line}

    
    \includegraphics[width=\columnwidth]{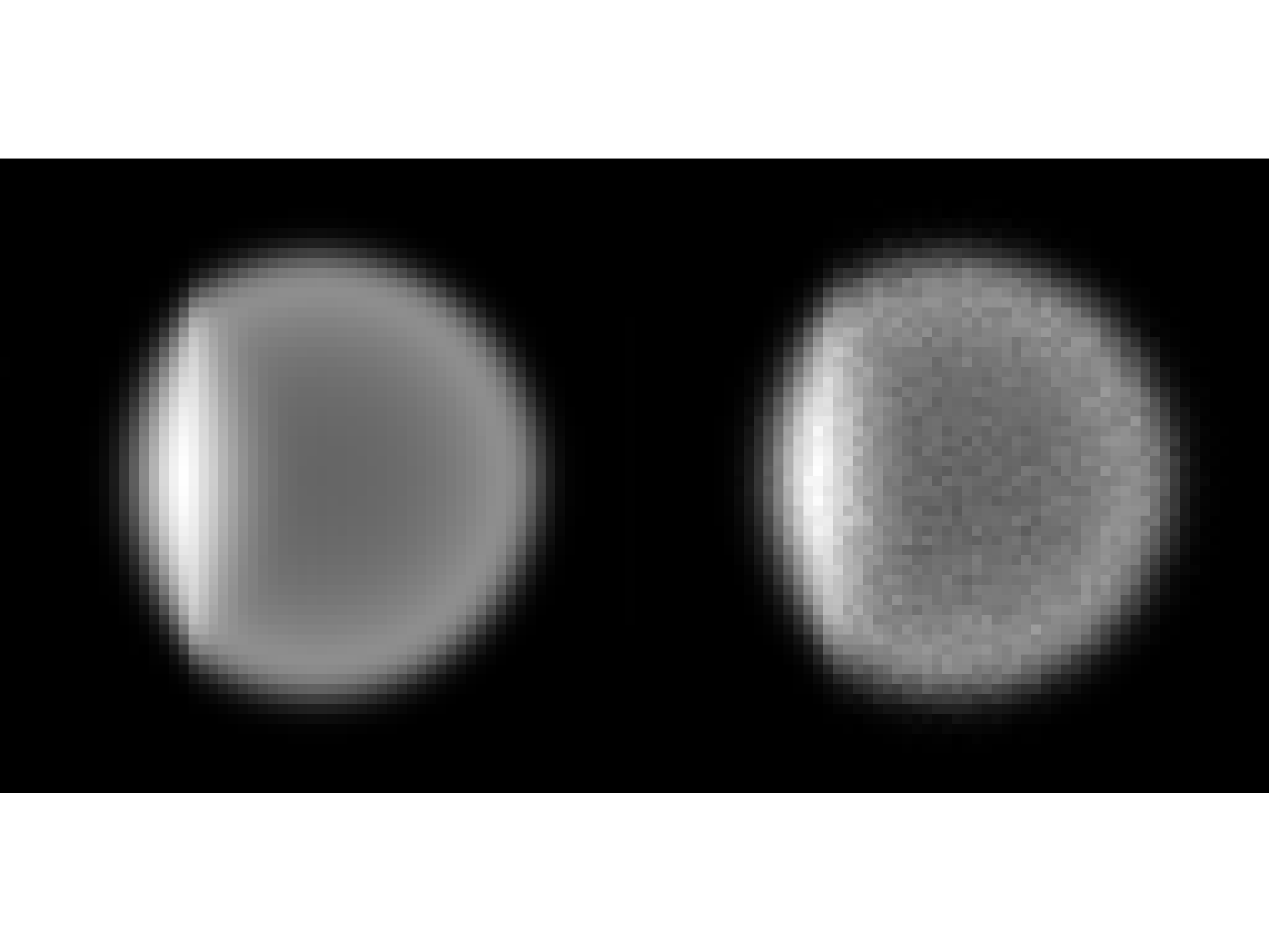}
    \caption{Fluence distribution in the domain at $z=25$mm and $t=65$ps, pseudo-spectral method (left) vs. Monte-Carlo (right).}
    \label{fig:inf_plane}
    
\end{figure}

The sharp onset of the scattering perturbation, and otherwise low scattering, provide an artificially challenging numerical scenario for the pseudo-spectral solver, promoting the propagation of higher angular harmonics and generating sharp features in the resultant fluence field.

A time-step of $0.1$ps was chosen by inspection, and $400$ steps were taken such that the initial radiance distribution, consisting of an Gaussian of isotropic angular profile, with $1/e$ diameter $1.5$mm, has time to propagate close the boundary of the domain. The simulation took $70$s to complete.

The same parameters were also employed in a Monte-Carlo solver of the RTE, in which $5\times10^8$ packets were propagated, with a kernel run-time of $269$s.

Figure \ref{fig:inf_line} demonstrates the evolution of the radiance field during propagation over a line through the domain. Figure \ref{fig:inf_plane} shows the resultant fluence through the domain after $60$ps, when the simulation was terminated. The initial field spreads outwards through the region of low-scattering close to the speed of light, forming a circular disc in the plane. As the radiance encounters the region of increased scattering a characteristic region of back-scattering causes a peak at the interface, and the further transport is inhibited and the field becomes diffuse.

An exact comparison is challenging since the time-domain Monte-Carlo solution is inherently averaged spatially and temporally (whereas the pseudospectral method is a collocation technique, with an exact solution  at points in space and time). Further reducing the spatial and temporal discretisation of the MC solution leads to an unacceptable reduction in SNR, highlighting the value of the proposed technique. Nonetheless, excellent agreement is found between the two solutions.

\subsection{Steady-state internal fluence with PML}

The effect of the PML can be evaluated by considering the steady-state fluence in the domain resulting from a given source distribution. 

The real domain consists of a cuboid of $35.5$mm on a side with homogeneous optical parameters $\mua = 0.01$mm$^{-1}$, $\mus=10.0$mm$^{-1}$, and anisotropy factor $g=0.95$. The real domain is embedded into a computation domain $50$mm on a side with a PML attenuation term following a half-cosine profile rising from $0.01$mm$^{-1}$ at the end of the real domain to $1.0$mm$^{-1}$ at the boundary. The domain is discretised into a $64\times64\times64$ grid of points, and the pseudo-spectral method is used to solve the system under a 7-degree spherical harmonic ($P_{7}$) approximation. 

\begin{figure}[t!]

    \centering
    
    \input{./figures/pml_x.tex}
    \caption{\label{fig:pml_x}Steady-state fluence distribution in the domain across the $x$ dimension at three points in $y$, at $z=0$mm, pseudo-spectral method (solid) vs. Monte-Carlo (dots).}
   
   \vskip 2em
   
    \input{./figures/pml_y.tex}
    \caption{\label{fig:pml_y}Steady-state fluence distribution in the domain across the $y$ dimension at three points in $x$, at $z=0$mm, pseudo-spectral method (solid) vs. Monte-Carlo (dots).}
    
\end{figure}

A time step of $0.04$ps was chosen by inspection, and $25\times10^3$ time steps were integrated to ensure the domain had reached steady state conditions. The source term was an isotropic Gaussian distribution, with $1/e$ diameter $1.5$mm, offset from the centre of the domain by $8$mm in the $x$-axis. The simulation took 223s to complete.

The same parameters were also employed in a Monte-Carlo solver of the RTE, in which only the real domain was simulated and a matched boundary condition employed. The MC simulation employed $1\times10^8$ packets, with a kernel run-time of $16.1$s.

Figures \ref{fig:pml_x} and \ref{fig:pml_y} demonstrate the resultant steady state fluence for both the pseudo-spectral and Monte-Carlo techniques. In each case the characteristic logarithmic decay of the fluence is evident away from the source and boundary, with an increased decay towards the boundary. The index-matched boundary conditions imply a non-zero fluence at the boundary the domain, evident in each simulation.

Excellent agreement is shown between the pseudo-spectral method and the Monte-Carlo solution.

\section{Discussion \& Conclusions}

We have introduced a new technique to solve the radiative transport equation in arbitrary heterogeneous media. The technique has been validated against a stochastic Monte-Carlo solver, demonstrating excellent agreement.

The use of a pseudo-spectral method allows a compact representation of the radiance field with sparser sampling than required in a finite-difference or finite-element scheme. The method is inherently parallelisable: whilst our implementation only uses a single GPU, the cuFFT library and CUDA kernels implement the algorithm can be readily extended to run across a number of devices with a minimum of inter-device communication. Our implementation is limited by the performance of the FFT, and thus the method demonstrates demonstrates $\mathcal{O}(N\log N)$ performance scaling.

The pseudospectral technique is particularly suited to biomedical imaging where inherent scattering leads to (angularly) smooth solutions which can be simulated with modest angular discretisations (e.g., $N_l \leq 7$). As we have demonstrated, the technique can readily support propagation in low-scattering regions when a higher degree approximation is employed. The absence of noise and deterministic performance is such that the technique could be employed as part of a conventional image reconstruction method in modalities such as diffuse optical tomography, quantitative photoacoustic tomography, and ultrasound modulated optical tomography.

The simplicity and flexibility of the technique suggests numerous avenues for further research. A particularly interesting avenue of investigation is the optimisation of the pseudo-spectral method for the computation of steady-state solutions. We noted earlier that the use of a Krylov solver to form such solutions directly from the integrated transport operator was an alternative to the relatively inefficient numerical integration of the time-domain solution. At present the computational performance of this approach is limited as Krylov methods are typically implemented on the CPU, necessitating significant overhead in memory transfers between the GPU and host. Overcoming this  would yield an attractive technique. We have also successfully employed exponential integrators to implement arbitrary time steps in homogenous media, allowing for more efficient numerical integration. 

\section{Acknowledgements}

S. Powell acknowledges support from RAEng Fellowship RF1516\textbackslash 15\textbackslash 33. S. Powell and S. R. Arridge acknowledge support from EPSRC grant EP/N032055/1. The authors declare no relevant conflicts of interests in this work.

\bibliography{psrte}

\end{document}